\documentclass[final,5p,times,twocolumn]{elsarticle}
\usepackage{amssymb,color}

\usepackage{hyperref}

\begin{document}
\begin{frontmatter}
\title{Diagnosing the cosmic coincidence problem and its evolution with recent observations}

\author{Jie Zheng$^1$,Yun Chen$^{2,3,*}$,Tengpeng Xu$^{2,3}$ and Zong-Hong Zhu$^{1,4}$}

\address{$^{1}$ Gravitational Wave and Cosmology Laboratory, Department of Astronomy, Beijing Normal University, Beijing 100875, China}
\address{$^{2}$ Key Laboratory for Computational Astrophysics, National Astronomical Observatories, Chinese Academy of Sciences, Beijing 100101, China \\
chenyun@bao.ac.cn}
\address{$^{3}$ College of Astronomy and Space Sciences, University of Chinese Academy of Sciences,Beijing 100049, China}
\address{$^{4}$ School of Physics and Technology, Wuhan University, Wuhan 430072, China}


\begin{abstract}
In the framework of a phenomenological cosmological model with the assumption of $\rho_{X}  \propto \rho_{m} a^{\xi}$ ($\rho_{X}$ and $\rho_{m} $ are the energy densities of dark energy and matter, respectively.), we intend to diagnose the cosmic coincidence problem by using the recent samples of Type Ia supernovae (SNe Ia), baryon acoustic oscillation (BAO) and cosmic microwave background (CMB).
$\xi$ is a key parameter to characterize the severity of the coincidence problem, wherein $\xi=3$ and $0$ correspond to the $\Lambda$CDM scenario 
and the self-similar solution without the coincidence problem, respectively.
The case of $\xi = Constant$ has been investigated in the previous studies, while we further consider the case of $\xi(z) = \xi_{0} + \xi_{z}*\frac{z}{1+z}$ to explore the possible evolution.
A joint analysis of the Pantheon SNe Ia sample with the recent BAO and CMB data figures out that $\xi=3.28\pm0.15$ in the case of $\xi = Constant$ at $68\%$ confidence level (CL), in addition, $\xi_{0} = 2.78_{-1.01}^{+0.28}$ and $\xi_{z} = 0.93_{-0.91}^{+1.56}$ in the case of $\xi(z)$ at $68\%$ CL.
It implies that the temporal evolution of the scaling parameter $\xi$ is accepted by the joint sample at $68\%$ CL; however, the joint sample also cannot distinguish whether the scaling parameter $\xi$ is variable or not at 95\% CL. Moreover, the $\Lambda$CDM scenario is accepted by the joint sample at $95\%$ CL in both cases, and the coincidence problem still exists.  In addition, we apply the Bayesian evidence to compare the models with the analysis of the joint sample, it turns out that the $\Lambda$CDM scenario is most supported by the joint sample; 
furthermore, the joint sample prefers the scenario with a constant $\xi$ to the one with a variable $\xi(z)$.
\end{abstract}


\end{frontmatter}



\section{Introduction}
\label{sec:intro}
The existence of an exotic form of energy with negative
pressure, dubbed ``dark energy'', is one of the most widely involved mechanism to explain the accelerating universe. The most popular dark energy models mainly include the $\Lambda$CDM model and the scalar-field dark energy model. Moreover, the $\Lambda$CDM model is preferred by most observations, though a small number of observations display a slight deviation \citep{Bull_et_al_2016, Bullock_and_Boylan-Kolchin_2017}. However, on the theoretical level the $\Lambda$CDM model is embarrassed by the well-known cosmological constant problems \citep{Weinberg_1989,Carroll_1992}, i.e., the  ``coincidence'' and ``fine-tuning'' problems. The ``coincidence problem'' states that why the present epoch is so special that the energy density of dark energy is in the same order of
magnitude as that of the matter only at this period. Several possible approaches have been adopted to explain or alleviate the coincidence problem, mainly including the anthropic principle \citep{Weinberg_2000,Vilenkin_2001,Garriga_et_al_2000,Garriga_and_Vilenkin_2001}, the slow evolving and spatially homogeneous scalar field with the ``tracker'' properties (see, for instance, \citep{Copeland_et_al_2006} for review), and the interaction between the dark energy and dark matter \citep{Amendola_2000,Caldera-Cabral_et_al_2009}.

In this work, we choose to explore the coincidence problem in a different perspective. A phenomenological model with minimal underlying theoretical assumptions is adopted, where the ratio of the energy densities of dark energy and matter is parameterized as $\rho_{X}  \propto \rho_{m} a^{\xi}$ \citep{Dalal_et_al_2001, Chen_et_al_2010}. This model originates from two special cases, i.e., $\rho_{X}  \propto \rho_{m} a^{3}$ for the $\Lambda$CDM model and $\rho_{X}  \propto \rho_{m} a^{0}$ for the self-similar solution without the coincidence problem,  where $\xi = 3$ and $0$, respectively.
The estimate value of $\xi$ obtained from the observational data can apparently reveal the severity of the coincidence problem. In addition, the standard cosmology without interaction between dark energy and dark matter is
characterized by $\xi + 3\omega_{X}= 0$, and  $\xi + 3\omega_{X} \neq 0$ indicates the non-standard cosmology. Furthermore,
any solution with a scaling parameter $0<{\xi}<3$ makes the coincidence problem less severe \citep{Pavon_et_al_2004}.

Besides the case of $\xi = Constant$ which has been studied in the previous works, we also explore the possible evolution of $\xi$ with the parametrization  $\xi(z) = \xi_{0} + \xi_{z}*\frac{z}{1+z}$.  The previous studies have conducted observational constraints on the scenario of $\xi = Constant$ with several different cosmological probes (see, for instance, \citep{Pavon_et_al_2004,Guo_et_al_2007,Chen_et_al_2010,Cao_et_al_2011,Zhang_et_al_2014}), including the SNe Ia, CMB, BAO, Hubble parameter $H(z)$ versus redshift and Sandage-Loeb test data sets.
In this work, by considering the cases of $\xi = Constant$ and $\xi(z) = \xi_{0} + \xi_{z}*\frac{z}{1+z}$, we explore the cosmic coincidence problem and its possible evolution with the recent observations, including the SNe Ia data from the Pantheon sample \citep{Scolnic_et_al_2018}, the CMB power spectrum data from the Planck 2018 final analysis \citep{Aghanim_et_al_2018}, and the BAO data from the measurements of  6dFGS survey\citep[]{Beutler2011}, SDSS DR7 MGS\citep[]{Ross2015the}, and BOSS DR12\citep[]{Alam2017}. 

This paper is organized as follows. In Section 2, we briefly introduce the phenomenological model under consideration. The Section 3 presents the observational data adopted in this work. The results from observational constraints and the corresponding analyses are displayed in Section 4. In the last section, we summarize the main conclusions.

\section{Phenomenological model: basic equations}

The model under consideration is characterized with a phenomenological form for the ratio of the dark energy and matter densities \citep{Dalal_et_al_2001,Chen_et_al_2010},
\begin{equation}
\rho_{X} \propto \rho_{m} a^{\xi}, \qquad or \qquad \Omega_{X} \propto \Omega_{m} a^{\xi},
\end{equation}
where $\Omega_{X}$ and $\Omega_{m}$ are the fractions of the energy density of the universe contributed from dark energy and matter, respectively. The scaling parameter ${\xi}$ can be constrained from observational data and used to reveal the severity of the coincidence problem.

Considering a flat FLRW universe with $\Omega_{X}+\Omega_{m}=1$, we can obtain
\begin{equation}
\Omega_{X} =\frac{\Omega_{X,0} a^{\xi}}{1-\Omega_{X,0}\left(1-a^{\xi}\right)},
\end{equation}
where $\Omega_{X,0}=\Omega_{X}(z=0)$. According to the energy conservation equation, we have
\begin{equation}
 \frac{d \rho_{\mathrm{tot}}}{d a}+\frac{3}{a}\left(1+\omega_{X} \Omega_{X}\right) \rho_{\mathrm{tot}}=0,
\label{eq:equation2}
\end{equation}
where $\rho_{\mathrm{tot}}=\rho_{m}+\rho_{X}$ is the total energy density, $\omega_{X}$ specifies the equation of state of the dark energy. Meanwhile, the Eq.(\ref{eq:equation2}) can be rewritten as
\begin{equation}
\frac{d \rho_{m}}{d a}+\frac{3}{a} \rho_{m} = -\left[\frac{d \rho_{X}}{da}+\frac{3}{a}\left(1+\omega_{X}\right) \rho_{X}\right] = Q,
\end{equation}
where $Q = -(\xi+3\omega_X)\rho_m \kappa a^{\xi-1}/(1+\kappa a^{\xi})$ and $\kappa = \rho_X/(\rho_m a^{\xi})$, and the interaction term
$Q = 0\;(\neq 0)$ denotes the cosmology without (with) interaction between dark energy and matter.

Based on Eq.(\ref{eq:equation2}), we can work out
\begin{equation}
\frac{\rho_{\mathrm{tot}}}{\rho_{0}}=\exp \left(\int_{a}^{1} \frac{d a}{a} 3\left(1+\omega_{X} \Omega_{X}\right)\right).
\label{eq:equation3}
\end{equation}
Assuming $\omega_{X}$ as a constant, we can rewritten the above equation as
\begin{equation}
E^{2}(z)=\exp \left(\int_{a}^{1} \frac{d a}{a} 3\left(1+\omega_{X} \Omega_{X}\right)\right),
\label{eq:equation4}
\end{equation}
where $E^2(z)\equiv [H(z)/H_0]^2 = \rho_{\mathrm{tot}}/\rho_{0}$, and $E(z)$ is the dimensionless Hubble parameter. When $\xi = Constant$, we can solve Eq.(\ref{eq:equation4}) and get
\begin{equation}
E^{2}(z;\textbf{p})=a^{-3}\left(1-\Omega_{X,0}\left(1-a^{\xi}\right)\right)^{-3 \omega_{X} / \xi},
\label{eq:equation5}
\end{equation}
where the parameter set is $\textbf{p} \equiv \left(\Omega_{X,0}, \omega_{X}, \xi\right)$.
However, for a variable $\xi(z)$,
$$\xi(z) = \xi_{0} + \xi_{z}*\frac{z}{1+z}.$$
We cannot obtain the analytical solution of Eq.(\ref{eq:equation4}). Then we should solve it numerically with the parameter set
$\textbf{p }\equiv \left(\Omega_{X,0}, \omega_{X}, \xi_{0},\xi_{z}\right)$.

\section{Data sample}
The observational data sets used in our cosmological analyses are described as follows,  including the Pantheon SNe Ia sample, the CMB power spectrum data from the final Planck 2018 results, and the BAO data from the 6dFGS survey, the SDSS DR7 MGS, and the BOSS DR12 measurements.

\subsection{SNe Ia data set}
The SNe Ia as standard candles have been proved to be a kind of sensitive probe of cosmology (see, e.g. \citep{Branch_and_Miller_1993, Riess_Press_and Kirshner_1996, Filippenko_2005}).
The population of confirmed SNe Ia has a dramatic increase over the last two decades, in the mean time, the techniques for measuring the light curve parameters are also continually being improved to reduce the systematic uncertainties. At present,
the most popular techniques mainly include the SALT/SALT2 \citep{Guy_et_al_2005, Guy_et_al_2007} and SiFTO \citep{Conley_et_al_2008} models, which are two popular techniques at present and fit the light curves of SNe Ia by using the spectral template.

 The SNe Ia sample adopted in this work is the Pantheon sample \citep{Scolnic_et_al_2018}, which consists of 1048 SNe Ia (0.01 $\le z \le$ 2.3) combined from  Pan-STARRS1(PS1) Medium Deep Survey, SDSS, SNLS, various low-z and HST samples.
 In the Pantheon sample, the distances for each of these SNe Ia are determined after fitting their light-curves with the most up-to-date published version of SALT2 \citep{Betoule_et_al_2014}, then applying the BEAMS with Bias Corrections (BBC) method \citep{Kessler_and_Scolnic_2017} to determine the nuisance parameters and adding the distance bias corrections. The uniform analysis procedure conducted on the SNe Ia of Pantheon sample has significantly reduced the systematic uncertainties related to photometric calibration.

The observable given in the Pantheon sample can be deemed as a correction to the apparent magnitude (see Table A17 of \citep{Scolnic_et_al_2018}), i.e.,
\begin{eqnarray}
Y^{obs} &=& m_B+K \nonumber\\
        &=& \mu+M,
\label{eq:Y_obs}
\end{eqnarray}
where $\mu$ is the distance modulus, $m_B$ is the apparent B-band magnitude, $M$ is the absolute B-band magnitude of a fiducial SN Ia, and the correction term $K = \alpha x_1-\beta c+\Delta_M+\Delta_B$ includes the corrections related to four different sources (see \citep{Scolnic_et_al_2018} for more details). The corresponding theoretical (predicted) value is
\begin{eqnarray}
Y^{th}&=& 5\log(d_L)+25 +M \nonumber\\
&=&5\log[(1+z)D(z)]+ Y_0,
\label{eq:Y_th}
\end{eqnarray}
where the constant term $Y_0$ is written as $Y_0 = M+5log(\frac{cH_0^{-1}}{Mpc})+25$, and the luminosity distance $d_L$ and the normalized comoving distance $D(z)$ are related with each other through the following formula, i.e.,
\begin{equation}
d_L(z) = \frac{c(1 + z)}{H_0}D(z),
\end{equation}
where $c$ is the velocity of light.
In a flat universe,  $D(z)$ can be expressed as
\begin{equation}
D(z) = \int_0^z\frac{d\tilde{z}}{E(\tilde{z})},
\label{eq:D_z}
\end{equation}
where $E(z)$ can be worked out with Eq. (\ref{eq:equation4}) for the model under consideration.

The chi-square statistic for the Pantheon sample can be constructed as
\begin{equation}
\label{eq:chi2SNe}
\chi^2_{\textrm{SNe}}={\Delta \overrightarrow{Y}}^T\cdot\textbf{C}^{-1}\cdot{\Delta \overrightarrow{Y}},
\end{equation}
where the residual vector for the SNe Ia data in the Pantheon sample is $\Delta \overrightarrow{Y}_i = [Y^{obs}_i-Y^{th}(z_i; Y_0,\textbf{p})]$. The covariance matrix $\textbf{C}$ of the sample includes the contributions from both the statistical and systematic errors. The nuisance parameter, i.e., the constant term $Y_0$  is marginalized over with the analytical methodology presented in \citep{Giostri_et_al_2012}.

\subsection{BAO data set} 

\begin{table}
	\caption{The BAO data adopted in this work.}
	\centering
	\label{tab:table_BAO}
	\begin{tabular}{ccccc}
		\hline
		Survey & $z_{eff}$ & Measurement & Value & $\sigma$ \\
		\hline
		6dFGS & 0.106 & $r_{s}/D_{V}$ & 0.336 & 0.015 \\
	
		SDSS DR7 MGS & 0.15 & $D_{V}\left(r_{s,fid}/r_{s}\right)$ & 664 & 25 \\
	
		BOSS DR12 & 0.38 & $D_{M}\left(r_{s,fid}/r_{s}\right)$ & 1518 & -- \\
		
		 & 0.38 & $H(z)\left(r_{s} / r_{s,fid}\right)$ & 81.5 & --\\
		
		 & 0.51 & $D_{M}\left(r_{s,fid}/r_{s}\right)$ & 1977 & -- \\
		
		 & 0.51 & $H(z)\left(r_{s} / r_{s,fid}\right)$ & 90.4 & -- \\
		
		 & 0.61 & $D_{M}\left(r_{s,fid}/r_{s}\right)$ & 2283 & -- \\
		
		 & 0.61 & $H(z)\left(r_{s} / r_{s,fid}\right)$ & 97.3 & -- \\
		\hline
	\end{tabular}

\label{tab:baodata}
\end{table}
 The BAO data extracted from galaxy redshift surveys are also a kind of  powerful cosmological
probe \citep[]{eisenstein1998baryonic,eisenstein2005detection}.
The BAO data set used here is 
a combination of measurements from the
6dFGS at $z_{\rm{eff}}=0.106$ \citep[]{Beutler2011}, the SDSS DR7 Main
Galaxy Sample (MGS) at $z_{\rm{eff}}=0.15$ \citep[]{Ross2015the}, and the BOSS DR12 at $z_{\rm{eff}} = (0.38,0.51,0.61)$ \citep[]{Alam2017}. The corresponding measurements are listed in Table \ref{tab:baodata}. 

The observable quantities used in the measurements are expressed in terms of the transverse co-moving distance $D_M(z)$, the volume-averaged angular diameter distance $D_V(z)$, the Hubble rate $H(z)\equiv H_0E(z)$,  the sound horizon at the drag epoch $r_s$, and its fiducial value $r_{\rm{s,fid}}$. 
Following \citep{Ryan_Chen_Ratra_2019}, we use the fitting formula of \citep{eisenstein1998baryonic} to compute $r_s$, and $r_{\rm{s,fid}}$ is computed with the fiducial cosmology adopted in the paper in which the measurement is reported.
In a flat universe, the transverse co-moving distance $D_M(z)$ equals to the line-of-sight comoving distance $D_{C}(z)$, which is expressed as,
\begin{equation}
D_{C}(z) \equiv \frac{c}{H_{0}} D(z),
\label{eq:equationDC}
\end{equation}
and $c$ is the speed of light.
The volume-averaged angular diameter distance is \begin{equation}
D_{V}(z)=\left[\frac{c z}{H_{0}} \frac{D_{M}^{2}(z)}{E(z)}\right]^{1 / 3}.
\label{eq:equationDVz}
\end{equation}
We employ the BAO data set in the analysis with the chi-squared statistic
\begin{equation}
\chi_{\mathrm{BAO}}^{2}(p)=\left[\vec{A}_{\mathrm{th}}(p)-\vec{A}_{\mathrm{obs}}\right]^{T} C^{-1}\left[\vec{A}_{\mathrm{th}}(p)-\vec{A}_{\mathrm{obs}}\right],
\label{eq:chi2_BAO}
\end{equation}
where $C^{-1}$ is the inverse of the covariance matrix. The BOSS DR12 measurements listed in the last six lines of Table \ref{tab:baodata} are correlated, and the corresponding  covariance matrix is present in Eq.(20) of \citep{Ryan_Chen_Ratra_2019}, which is also available from SDSS website\footnote{\url{https://sdss3.org/science/boss_publications.php}}.

\subsection{CMB data set} Observations of the CMB spectra provide another kind of independent test of the existence of dark energy.
It is remarkable that the CMB power spectra from the WMAP  \citep{Hinshaw2013WMAP} and Planck projects \citep{Aghanim_et_al_2018} have provided strong constraints on cosmological parameters. Here, we use the combination of temperature and polarization CMB power spectra from the Planck 2018 release \citep{Aghanim_et_al_2018}, including the likelihoods at multipoles $\ell=2-2508$ in TT, $\ell=2-1996$ in EE, and $\ell=30-1996$ in TE. 
In practice, different algorithms have been used to estimate the CMB power spectrum, such as Commander\citep{Planck_2018_A4,Planck_2018_A5}, SimAll\citep{Planck_2018_A5} and Pilk\citep{{Aghanim_et_al_2018}}.
The ``Commander'' component-separation algorithm is used to estimate the power spectrum over the range  $\ell=2-29$ in TT. The ``SimAll'' approach is used to estimate the power spectrum over the range   $\ell=2-29$ in EE. The ``Pilk'' cross-half-mission likelihood \citep{Planck_2018_A5} is used to compute the CMB high-$\ell$ part for TT,TE,EE over the range $30 \leq \ell \leq  2508$ in TT and over the range $30 \leq \ell \leq 1996$ in TE and EE \footnote{For more details on the Planck CMB spectrum and likelihood code, see \url{https://wiki.cosmos.esa.int/planckpla/index.php/CMB_spectrum_\%26_Likelihood_Code} }. Hereafter, $\mathcal{L}_{Planck}$ denotes the likelihood of the Planck data described above.

\section{Analysis and Results}
\subsection{Observational constraints}
\begin{table*}
\caption{\label{tab:parameters}
The mean values with $68\%$ confidence limits for model parameters constrained from the Pantheon SNe sample, and from a joint sample of SNe, BAO and CMB data sets, respectively. The scenarios with $\xi = Constant$ and $\xi(z) = \xi_{0} + \xi_{z}*\frac{z}{1+z}$ are both considered.}
\centering
\resizebox{\textwidth}{!}{
\begin{tabular}{ccccccc}
\hline		
Model & Data set & $\Omega_{X,0}$ & $\omega_{X}$ & $\xi$ & $\xi_{0}$ & $\xi_{z}$ \\
\hline
$\xi = Constant$ & Pantheon & $0.75_{-0.08}^{+0.13}$ &$-0.96_{-0.14}^{+0.16}$ & $3.42_{-0.62}^{+1.22}$ & - & - \\

$\xi = Constant$ & Pantheon + BAO + CMB  & $0.67\pm0.01$ & $-1.12\pm0.04$ & $3.28\pm0.15$ &- &- \\

$\xi(z) = \xi_{0} + \xi_{z}*\frac{z}{1+z}$ & Pantheon & $0.69^{+0.04}_{-0.15}$ & $-1.01_{-0.32}^{+0.04}$ & - & $3.00_{-0.78}^{+0.09}$ & $0.94_{-1.02}^{+0.58}$  
\\
$\xi(z) = \xi_{0} + \xi_{z}*\frac{z}{1+z}$ & Pantheon + BAO + CMB & $0.69\pm 0.01$ & $-0.99_{-0.06}^{+0.03}$ & - & $2.78_{-1.01}^{+0.28}$ & $0.93_{-0.91}^{+1.56}$  \\
\hline
\end{tabular}}
\end{table*}

\begin{figure*}
	\includegraphics[width=1.9\columnwidth]{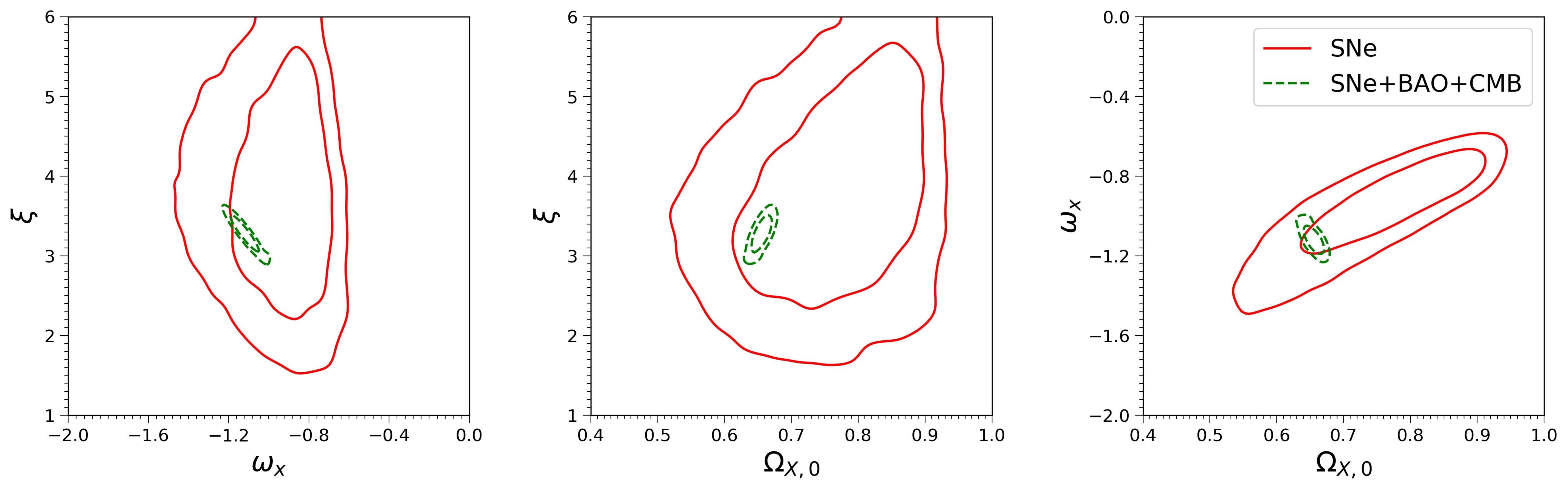}
    \caption{The 2D probability distributions of model parameters in the scenario of $\xi = Constant$, constrained from the Pantheon SNe sample (red solid lines), and from a joint sample of the SNe, BAO and CMB data (green dotted lines), respectively. The contours correspond to $68\%$ and $95\%$ CLs.}
    \label{fig:fig1}
\end{figure*}

\begin{figure*}
    \includegraphics[width=1.9\columnwidth]{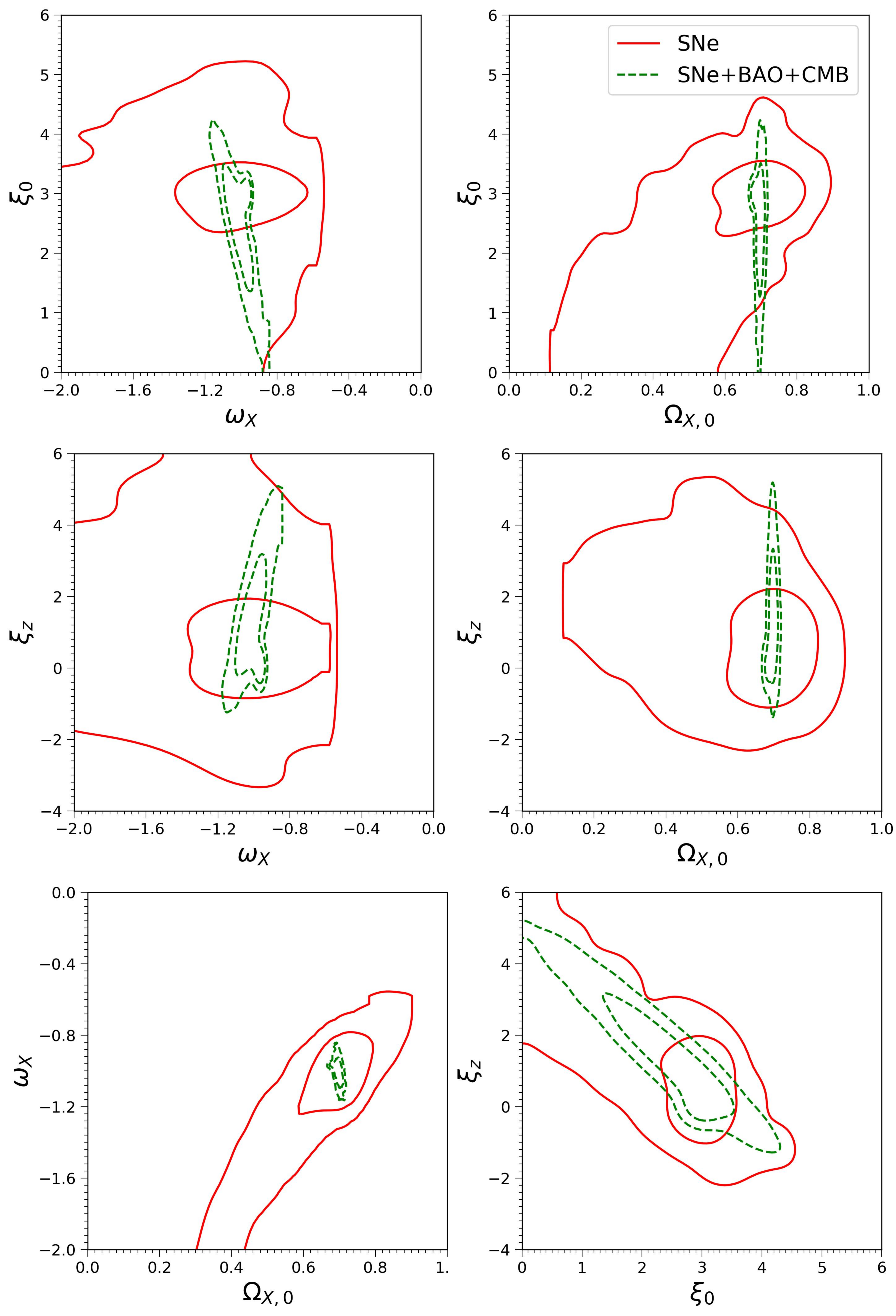} 
    \caption{The 2D contours of parameters in the scenario of $\xi(z) = \xi_{0} + \xi_{z}*\frac{z}{1+z}$. The implications of line styles are the same as those in Fig.\ref{fig:fig1}.}
    \label{fig:fig2}
\end{figure*}

In our analysis, the total likelihood for parameters is
\begin{equation}
\label{eq:LH_total}
\mathcal{L}(\mathbf{p})=\prod \mathcal{L}_{i},
\end{equation}
where $\mathcal{L}_{i}$ means the likelihood of each data set. In the case of using the combination of SNe Ia, BAO and CMB data sets, it takes,
\begin{equation}
\mathcal{L}_{tot}(\mathbf{p})=\mathcal{L}_{SNe} \mathcal{L}_{BAO} \mathcal{L}_{Planck}
\end{equation}
We derive the posterior probability distributions of parameters with Markov Chain Monte Carlo (MCMC) exploration using the May 2020 version of CosmoMC \citep{Lewis_et_al_2002}.
In the next following analysis, we consider two different treatment schemes for the scaling parameter $\xi$, i.e., $\xi = Constant$ and $\xi(z) = \xi_{0} + \xi_{z}*\frac{z}{1+z}$.

The case of $\xi = Constant$ has been widely studied in the literature \citep[see e.g.][]{Pavon_et_al_2004,Guo_et_al_2007,Chen_et_al_2010,Cao_et_al_2011,Zhang_et_al_2014}. Here we re-explore this scenario with the latest data sets. In addition, the case of $\xi(z)$ is taken into account to explore the possible evolution. We put observational constraints on the model parameters with the recent Pantheon SNe Ia sample, as well as with a combination of the SNe Ia, BAO and CMB data sets, respectively. We present the mean values with $68\%$ confidence limits for the parameters of interest in Table~\ref{tab:parameters} for both scenarios. In the scenario of $\xi = Constant$, the constraints on $\Omega_{X,0}$, $\omega_X$ and $\xi$ from the combining sample are much tighter than those from the single Pantheon SNe Ia sample.
The constraints on the parameters $(\Omega_{X,0}, \omega_X, \xi)$ from the Pantheon SNe sample are consistent with those from the ``Constitution Set'' SNe sample adopted in \citep{Chen_et_al_2010} at $68\%$ CL. However, the constraints on $(\Omega_{X,0}, \omega_X, \xi)$ from our combining sample are inconsistent with those from the joint SNe + BAO + CMB sample adopted in  \citep{Chen_et_al_2010} at $68\%$ CL, but they are consistent at $95\%$ CL.
Moreover, the $\Lambda$CDM scenario, i.e., $(\omega_X, \xi) = (-1, 3)$, is accepted by the Pantheon SNe sample at $68\%$ CL, however, it's ruled out by the combining sample at $99\%$ CL.  In the scenario of $\xi(z)$, the constraints on $\Omega_{X,0}$ and  $\omega_X$ from the combining sample are much tighter than those from the Pantheon SNe sample, but the constraint precisions on $\xi_0$ and $\xi_z$ from the combining sample do not have significant improvements compared with those from the single Pantheon SNe sample. The $\Lambda$CDM scenario, i.e., $(\omega_X, \xi_0, \xi_z) = (-1, 3, 0)$, is accepted by the Pantheon SNe sample but ruled by the combining sample at $68\%$ CL, nevertheless, it's accepted by the combining sample at 95\% CL. Then, we pay attention to the constraints on the parameter $\xi_z$ which indicates the degree of temporal evolution of the scaling parameter $\xi$. The mean values with $68\%$ confidence limits for the parameters $\xi_z$  are  $\xi_z =0.94^{+0.58}_{-1.02}$ from the Pantheon SNe sample and $\xi_z =0.93^{+1.56}_{-0.91}$ from the combining sample. It implies that the Pantheon SNe sample cannot distinguish between the evolving and non-evolving scenarios, and the combining sample supports the time-evolving scenario at $68\%$ CL. 

The two-dimensional (2D) contours for the model parameters of interest are presented in Fig.~\ref{fig:fig1} for the scenario of $\xi=Constant$ and in Fig.~\ref{fig:fig2} for the scenario of $\xi(z)$.  
From Fig.~\ref{fig:fig1}, one also can see that the constraints from the combining sample are much more restrictive than those from the Pantheon SNe sample, though there are degeneracies  between some parameters. The $\omega_X-\xi$ plane does not have significant degeneracy from the Pantheon SNe sample, but displays a negative correlation from the combining sample. The $\Omega_{X,0}-\xi$ plane demonstrates a positive correlation from both the single Pantheon SNe sample and the combining sample. Especially,
the $\Omega_{X,0}-\omega_X$ plane displays a positive correlation from the Pantheon sample, conversely, a negative correlation from the combining sample. From Fig.~\ref{fig:fig2}, one can find out that  
the contours constrained from the combining sample shrink significantly compared with those from the Pantheon SNe sample except for the last panel, i.e., the $\xi_0 - \xi_z$ plane.  It implies that the addition of the BAO and CMB data sets cannot greatly improve the constraint precisions on $\xi_0$ and $\xi_z$.

\subsection{Model selection statistics}
\begin{table}\caption{\label{tab:table_selection}We list the natural logarithm of the Bayesian evidences $\ln B_{i}$ and the Bayes factors $\ln B_{i,0}$ from the joint sample of SNe+BAO+CMB, where the subscript ``0'' denotes the $\Lambda$CDM model.}
\centering

	\begin{tabular}{ccc}
		\hline
		Model &  $\ln B_{i}$ & $\ln B_{i,0}$ \\
		\hline
		$\Lambda$CDM & -1940.80 & 0 \\
		
		$\rho_{X}  \propto \rho_{m} a^{\xi}$  with $\xi = Constant$ & -2022.22  & -81.42 \\
		
		$\rho_{X}  \propto \rho_{m} a^{\xi}$  with $\xi(z) = \xi_{0} + \xi_{z}*\frac{z}{1+z}$ & -2030.04 & -89.24 \\
		\hline
	\end{tabular}
\end{table}

In the framework of Bayes' theorem, the probability that the model $M_{i}$ is true can be estimated with 
\begin{equation}
P\left(M_{i} \mid D\right)=\frac{P\left(D \mid M_{i}\right) P\left(M_{i}\right)}{P(D)},
\end{equation}
where $P(M_{i} \mid D)$ is the posterior probability, $D$ denotes the observational data, $P(M_{i})$ is a prior probability in the model $M_{i}$, and  $P(D)$ is the normalization constant. In addition, $P(D \mid M_{i})$ is the so-called Bayesian evidence \citep{roberto2008,limitation2008}, which can be written as
\begin{equation}
P\left(D \mid M_{i}\right)=\int P\left(D \mid \bar{\theta}, M_{i}\right) P\left(\bar{\theta} \mid M_{i}\right) d \bar{\theta}, 
\end{equation}
where $P(D \mid \bar{\theta},M_{i})$ is the likelihood function under the model $M_i$, and $P(\bar{\theta} \mid M_{i})$ is the prior probability for parameter $\bar{\theta}$ under the model $M_i$. Hence, calculating the Bayesian evidence requires the evaluation of an integral over the entire likelihood function and the prior distributions of model parameters.
When comparing two models, e.g., $M_{i}$ versus $M_{j}$, the Bayes factor
\begin{equation}
B_{i j}=\frac{P\left(D \mid M_{i}\right)}{P\left(D \mid M_{j}\right)},
\end{equation}
 which is defined as the ratio of the Bayesian evidences of two models can be employed as a judgment criterion, where the Bayes factor $B_{ij}>1$ (i.e., $\ln B_{ij} > 0$) means that the observational data prefer $M_i$ to $M_j$, and $B_{ij}<1$ implies that $M_j$ is preferred \citep{bayesfactor}. 
 
To compare the phenomenological models under consideration with the $\Lambda$CDM model, we calculate the values of Bayesian evidence for each model, where the code 
$\textbf{MCEvidence}$ \citep{MCEvidence} which is a popular python package to compute the Bayesian evidence is adopted here, and the observational data correspond to the joint sample of SNe, BAO and CMB data. 
In Table \ref{tab:table_selection}, we show the natural logarithm of the Bayesian evidence for each model, $\ln B_i$, as well as the natural logarithm of the Bayes factor, $\ln B_{i0}$, where the subscript ``0'' denotes the $\Lambda$CDM model.
It turns out that the $\Lambda$CDM model is most supported by the joint sample, since $B_{1,0}$ and $B_{2,0}$ are both smaller than 1, where the subscripts ``1'' and ``2'' denote  the scenarios with a constant $\xi$ and a variable $\xi(z)$, respectively. In addition, $B_{1,2} =B_1/B_2$ is bigger than 1, so the scenario with a constant $\xi$ is more competitive than the one with a variable $\xi_z$.  

\section{Summary and conclusions} 
We have concentrated on a kind of phenomenological model of cosmology, where the assumption of $\rho_{X} \propto \rho_{m} a^{\xi}$ is adopted. As a key parameter, the scaling parameter $\xi$ reveals the severity of the coincidence problem, where the particular values $\xi = 3$ and $\xi = 0$ correspond to the $\Lambda$CDM scenario and the self-similar solution without coincidence problem, respectively. Besides the scheme of assuming $\xi = Constant$,  we have also considered the scenario with a variable $\xi(z) = \xi_{0} + \xi_{z}*\frac{z}{1+z}$ to explore the possible evolution. The observational constraints on the model parameters are conducted with both the single Pantheon SNe Ia sample and a joint sample of SNe, BAO and CMB data sets, where the CMB power spectrum data are from the Planck 2018 final analysis , and the BAO data are from the measurements of 6dFGS, SDSS DR7 MGS, and BOSS DR12. 

The main conclusions can be summarized as follows:
(i) In the case of $\xi = Constant$, the $\Lambda$CDM scenario, i.e., $(w_X, \xi) = (-1, 3)$, is accepted by the Pantheon SNe sample and by the joint sample at $68\%$ CL and $95\%$ CL, respectively. Moreover, in the case of a variable $\xi(z)$, the $\Lambda$CDM scenario, i.e., $(w_X, \xi_0, \xi_z) = (-1, 3, 0)$, is also accepted by the Pantheon SNe sample and by the joint sample at $68\%$ CL and $95\%$ CL, respectively. 
(ii) According to the observational constraints on the model parameters,  the Pantheon SNe sample cannot distinguish between the scenarios of a constant $\xi$ and a variable $\xi(z)$ at $95\%$ CL because of $\xi_z\in [-2.60, 5.23]$ at 95\% CL; moreover, the joint sample also cannot distinguish whether the scaling parameter $\xi$ is variable or not at 95\% CL because of $\xi_z\in [-0.67, 4.40]$ at 95\% CL. 
(iii) According to the Bayesian evidences calculated from the joint sample, we find out that the $\Lambda$CDM model is most supported by the joint sample; furthermore, the joint sample prefers the scenario with a constant $\xi$ to the one with a variable $\xi(z)$.
(iv) The inclusion of the BAO and CMB data sets just can provide very limited improvements on constraining $\xi_0$ and $\xi_z$ in the scenario of $\xi(z)$, but it has significantly reduced the allowed regions of other parameters. Thus, to diagnose the evolution of the scaling parameter $\xi$ more robustly, it seems to be quite necessary to 
explore other probes which can supply more efficient improvements on constraining $\xi_0$ and $\xi_z$.   

\section{acknowledgments}
This work has been supported by the National Natural Science Foundation of China
(Nos. 11988101, 11633001, 11920101003, 11703034, 11773032 and 11573031), the Strategic Priority Research
Program of the Chinese Academy of Sciences (No. XDB23000000), the Interdiscipline Research Funds
of Beijing Normal University, and the NAOC Nebula Talents Program.

\paragraph{Note added.} The data underlying this article will be shared on reasonable request to the corresponding author.

\end{document}